\documentclass{article}
\usepackage{spconf,amsmath,graphicx}
\usepackage{multirow}
\usepackage{booktabs}       

\title{Fast Contextual Adaptation with Neural Associative Memory for On-Device Personalized Speech Recognition}
\name{
\begin{tabular}{c}
Tsendsuren Munkhdalai, Khe Chai Sim, Angad Chandorkar, Fan Gao, Mason Chua,\\
Trevor Strohman, Fran{\c{c}}oise Beaufays
\end{tabular}
}
\address{Google, USA \\
\fontsize{9}{9}\selectfont\ttfamily\upshape
\{tsendsuren, khechai, angadc, fgao, aftran, strohman, fsb\}@google.com}
\begin{document}
\ninept
\maketitle
\begin{abstract}
Fast contextual adaptation has shown to be effective in improving Automatic Speech Recognition (ASR) of rare words and when combined with an on-device personalized training, it can yield an even better recognition result. However, the traditional re-scoring approaches based on an external language model is prone to diverge during the personalized training. In this work, we introduce a model-based end-to-end contextual adaptation approach that is decoder-agnostic and amenable to on-device personalization. Our on-device simulation experiments demonstrate that the proposed approach outperforms the traditional re-scoring technique by 12\% relative WER and 15.7\% entity mention specific F1-score in a continues personalization scenario.
\end{abstract}
\begin{keywords}
speech recognition, on-device learning, fast contextual adaptation
\end{keywords}

\section{Introduction}
\label{sec:introduction}
ASR on mobile devices has reached a good performance and many real world applications now rely on a voice-based interface~\cite{schalkwyk2019}. However, recognizing rare and Out-Of-Vocablary (OOV) words such as entity mentions and domain specific terminologies still presents a significant challenge to the state-of-the-art ASR systems, due to the long-tail nature of the word distribution~\cite{czarnowska2019don}. 

Personalization of ASR models has been successful in improving recognition of user specific long-tail inputs~\cite{sim2019personalization,sim21_interspeech}. The on-device model personalization needs additional supervision and training. On the other hand, fast ASR model adaptation with contextual information can immediately boost the performance on OOV words and domain specific terms without requiring additional supervised data and training. It can also help disambiguating phonetically similar words with different semantics by processing a rich contextual information. Therefore, the fast contextual adaptation complements the model personalization and when both approaches are combined, they offer an efficient way to address the current limitations relating to the long-tail distribution.

There are two generic techniques to extend the ASR model with the fast contextual adaptation ability. A fast adaptation component can be either added as an external module on top of the recognition output or built jointly into the neural ASR model in an end-to-end fashion. The former approach makes use of an external Language Model (LM) and interpolates the ASR and the LM distributions to re-score the output prediction during decoding. Here the LM is separately built to encode the contextual information. The tradition Finite State Transducer (FST) based contextual biasing is common in achieving this~\cite{aleksic2015improved,mcgraw2016personalized}.



\begin{figure*}[t]

\begin{minipage}[b]{.33\linewidth}
  \centering
  \centerline{\includegraphics[width=3.5cm]{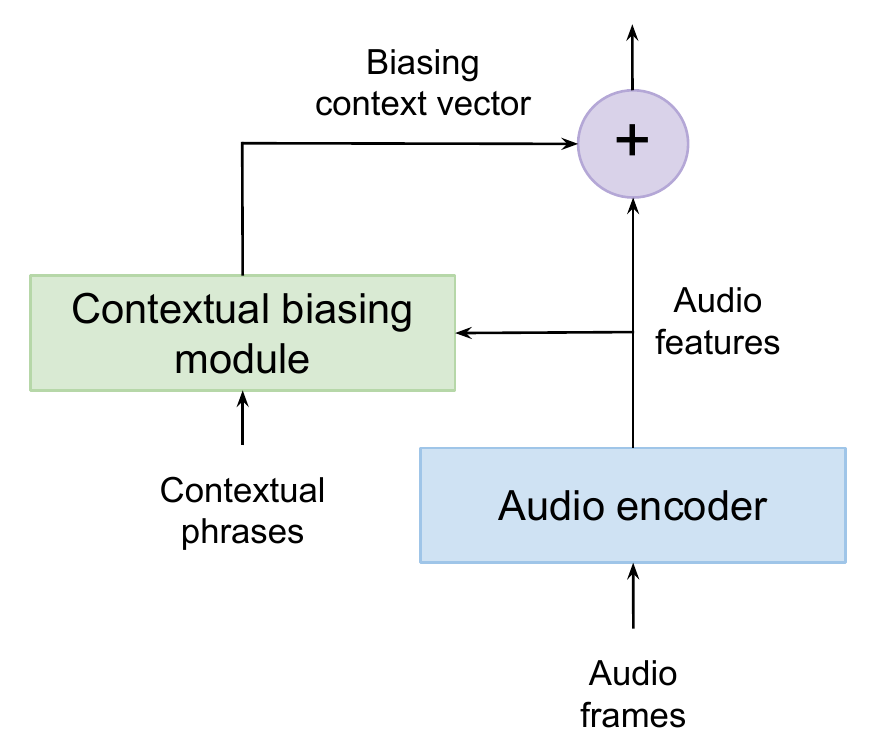}}
  \centerline{(a)}\medskip
  \label{fig:enc}
\end{minipage}
\hfill
\begin{minipage}[b]{.7\linewidth}
  \centering
  \centerline{\includegraphics[width=11cm]{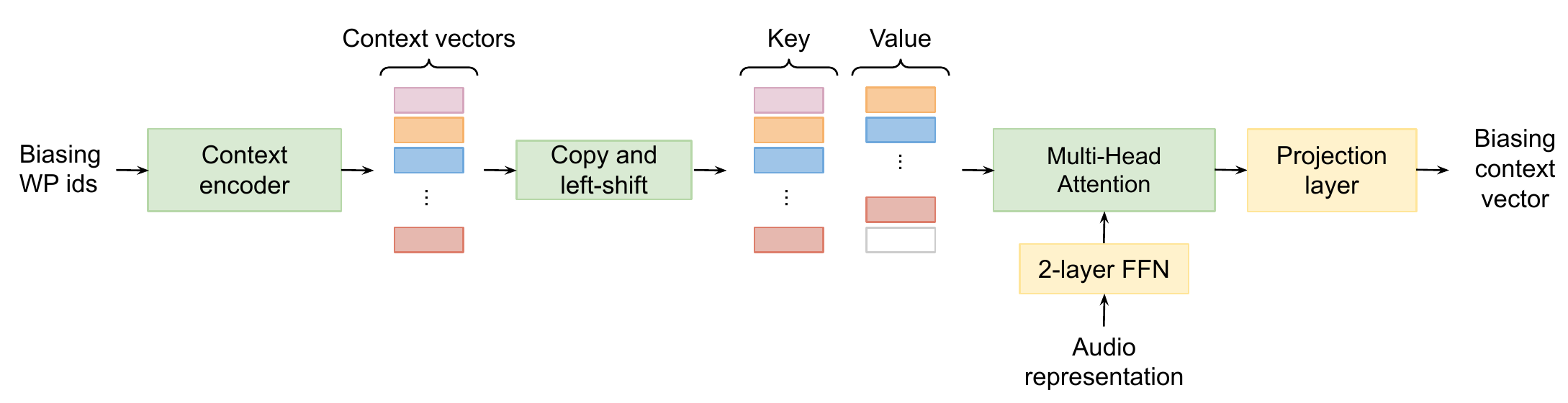}}
  \centerline{(b)}\medskip
  \label{fig:biaser}
\end{minipage}
\caption{Overall architecture of (a) our audio encoder and (b) the proposed contextual biasing module.}
\label{fig:encoder}
\end{figure*}

Unlike the traditional re-scoring methods, the end-to-end approaches jointly train a neural contextual module as a part of the ASR model so that the entire model is optimized to acquire the fast adaptation behavior~\cite{munkhdalai2017meta,munkhdalai2020sparse} during training. Along this line, \cite{pundak2018deep} extended LAS~\cite{chan2016listen} with a contextual module and introduced the Contextual-LAS (CLAS) model. The CLAS module takes in a set of contextual phrases and constructs their embeddings. It then uses an additive attention mechanism~\cite{bahdanau2014neural} to retrieve the relevant context given a partially decoded transcript hypothesis. Finally, the context is combined with the other audio and language features and passed to the decoder layers. A follow-up work also demonstrated that the performance of the CLAS model can be improved by training on carefully sampled hard negative contextual (i.e. anti-context) phrases~\cite{alon2019contextual}. More recently, \cite{jain2020contextual} proposed a similar contextual biasing module for the RNN-T model~\cite{graves2012sequence,graves2013sequence} in an open domain setting. Similar to CLAS, they perform a soft-matching between the partially generated hypothesis and the contextual phrases with the additive attention for relevant context.

The end-to-end approaches are robust, specially in a dynamic adaptation scenario where the ASR model undergoes a continuous on-device personalization as the different parts of the models are tightly integrated and optimized together whereas the FST biasing can diverge, requiring re-calibration of the independent components that are trained separately. In this work, we introduce an end-to-end fast contextual biasing approach for personalized speech recognition. Our technique consists of a novel Neural Associative Memory (NAM) and a multi-head attention mechanism for efficient contextual adaptation. The novel associative memory stores a chain of transitions between the sub-sequences in each contextual phrase and the multi-head attention mechanism computes the relevant contextual representation from the memory content. 

In our contextual module, we utilize audio features by leveraging multi-modal learning~\cite{pmlr-v139-ramesh21a} as shown in Figure~\ref{fig:encoder}a. Therefore, unlike the previous CLAS model and the other similar approaches that require partially decoded text to work, our proposed method relies only on the audio encoder features making it agnostic from the choice of a decoder. The experimental results demonstrate that our contextual adaptation technique outperforms the previous model-based end-to-end baseline (i.e. CLAS-\textit{encoder}) and the FST-based approach and when further combined with the continuous personalized training, it achieves a consistent improvement with each training round. We also analyze our model by performing an ablation study and comparing with the traditional FST biasing on different contextual regimes.

\section{Methods}
\label{sec:methods}
Figure~\ref{fig:encoder}a illustrates the audio encoder architecture of our NAM model for on-device personalized speech recognition. The encoder consists of a contextual biasing model and an audio encoder. In addition to the standard audio input, it has another external input for processing the contextual information. In this work, we focus on the contextual information provided in the form of text although our approach can readily be extended to other form of contextual information such as image and video.

The model starts by processing the input audio frames and extracting corresponding audio features with an audio encoder. We then provide a set of contextual phrases to the contextual biasing module and compute a relevant context for each audio feature vector. Finally, the context vectors are applied as a shift to the original audio features to produce the contextualized audio representations. 

\subsection{Contextual Biasing Module}
\label{sec:biaser}

Figure~\ref{fig:encoder}b shows the overall architecture of the proposed contextual biasing model. This module takes in a set of contextual phrases and a sequence of audio features $h_t \in {\rm I\!R}^{e}$ and computes the relevant context representations. First, the word-piece (WP) tokens for the entire contextual phrases are mapped to dense context vectors with a text context encoder. We then build a key-value associative memory and store the resulting representations. Finally, given an audio representation, a relevant biasing context vector is constructed from the memory by using a Multi-Head Attention (MHA) mechanism and passed to the projection layer. 

\subsubsection{Context Encoder}
\label{sec:encoder}
The context encoder performs a bi-directional contextual embeddings of each contextual phrase. Given $B$ number of contextual phrases and their WP ids denoted as $\{w_{u,i}\}^B_{i=1}$ where $u$ is the WP index, the encoder computes their corresponding context vectors $\{x_{u,i}\}^B_{i=1}$ where $x_{u,i} \in {\rm I\!R}^{d}$. Any bi-directional text embedding methods including BiLSTM~\cite{hochreiter1997long} can be used as the encoder. In this work, we parameterize the context encoder with TransformerXL~\cite{dai2019transformer} without causal attention masks.

\subsubsection{Neural Associative Memory}
\label{sec:kv-am}
The CLAS and its related techniques~\cite{pundak2018deep,jain2020contextual} construct a dense representation for each contextual phrase for a coarse-grained global biasing. They do not explicitly model the transition probability between the WPs tokens. Modelling this transition can be useful when biasing for personalized entity names and proper nouns that are rare and unseen during training as it allows us to recover an expected next token by using its preceding sub-sequence. Therefore, this work introduces a more fine-grained biasing technique that operate at the WP-level. 

Our goal is to learn and utilize what WP token is likely to follow given a sequence of preceding WPs. This can be achieved by the standard LM objective that models the conditional probability of a current token $w_{u,i}$ given its preceding tokens $w_{u,<i-1}$ as $p_{LM}(w_{u,i}|w_{u,<i-1})$. However, learning the conditional distribution in an end-to-end model-based fashion is a slow process and it is computationally infeasible to do this on the fly for biasing contexts that change dynamically. On the other hand, an associative memory~\cite{hopfield1982neural} can reconstruct a previously stored pattern by using its partial or corrupted variant~\cite{munkhdalai2017neural,schlag2020learning}. In this work, we propose to amortize this process by using a fast associative memory mechanism. In our contextual biasing module, whenever the biasing context changes with new contextual phrases, we update our memory to learn their conditional transition on the fly and then during decoding, we retrieve the next WP representation from the memory to inform the decoder about an expected output at the current step.

Concretely, once the context representations $\{x_{u,i}\}^B_{i=1}$ are constructed for the entire input set, we build an  associative memory to store and retrieve the relevant biasing contexts. As shown in Figure~\ref{fig:encoder}b, The memory stores the associative transition between the WP sub-sequences of the same phrase. 
In the associative memory, every WP token in each phrase is used as a key to map to the next WP token (left-shifted). Therefore, a memory item $l$ of a key-value pair $(k_{l}, v_{l})$ is built from the two successive context embeddings $x_{u-1,i}$ and $x_{u,i}$. In our memory, all the phrases are stored in a flat structure and there are a total of $B \times U$ number of key-value pairs. Here $U$ is the maximum length of the phrases and we perform padding and masking for shorter phrases. 

This creates a chain of transitions in our memory and allows us to easily recover the current token $x_{u}$ from the memory by using its previous element $x_{u-1}$ to traverse a contextual phrase in a fine-grained manner. To handle the cases where there is no relevant biasing context given an audio representation, we make use of a special slot in the memory with learned key and value vectors.

\subsubsection{Multi-Head Attention}
\label{sec:roe}
In order to retrieve relevant context from the representations of the contextual phrases, the previous approaches~\cite{alon2019contextual} rely on the original additive attention technique~\cite{bahdanau2014neural}. However, there are two main limitations related to this. As the attention focus is limited only on a single slot among the entire memory entries, the model is not able to aggregate multiple representations efficiently with a single forward-pass. Second, the standard single-head attention is prone to optimization difficulties~\cite{snell2021approximating}, which can limit its applications to harder tasks. The MHA mechanism improves by using multiple parallel heads in a single pass~\cite{vaswani2017attention} and thus it is suitable for our model. Specially, in our NAM model and unlike the other approaches, we compute the relevant biasing context by using the audio input rather than the partially decoded text for contextualized audio representations. Therefore, it is crucial for our attention mechanism be able to learn across multi-modalities. 

\subsection{Training}
\label{sec:training}
Our contextual biasing module can extend any neural ASR system with encoder-decoder architecture for fast contextual adaptation. We incorporate our biasing module into an existing ASR model that is already trained and then continue to train them jointly. Training converges quickly since most model parameters are already trained. This approach is more efficient than training the entire model from scratch. 
Since our contextual biasing approach only modifies the encoder outputs for fast adaptation, it is decoder-agnostic and thus can be used along with different ASR decoding approaches and training losses such as RNN-T~\cite{graves2012sequence,graves2013sequence}, offering a greater flexibility than the previous CLAS related approaches. 

In order for the model to learn to readily exploit the biasing context, a proper credit assignment is forced during training. Following~\cite{pundak2018deep}, we select an n-gram phrase from each transcript as the correct biasing context for that input and add this phrase to the entire context set with a probability $p$. We then annotate the transcript by marking each occurrence of the selected n-gram with a special token $<$/bias$>$. Therefore, the model must predict the special $<$/bias$>$ token by constructing the correct biasing context from the entire set. 

\section{Experiments}
\label{sec:experiments}

\begin{table}[t]
    \centering
    \caption{Results on {\em Wiki-Names} test set. For the FST baseline, we reported multiple results with different re-scoring weights.}
    \label{tab:adaptation}
    \begin{tabular}{c c c c c}
        \toprule
        \multirow{1}*{Model} & WER & Precision & Recall & F1-score \\
        \toprule
        \multirow{1}*{None}
        &    36.0    &  \bf 98.1    &    12.3    &    21.8  \\
        \midrule
        \multirow{4}*{FST}
        &   32.5 & 95.0 &	37.3 &	53.5\\
        &   29.7 & 92.5 &	58.6 &	71.7\\
        &   29.4 & 73.9 &	70.4 &	72.1\\
        &   46.8 & 4.7 &	\bf 77.2 &	8.8\\
        \midrule
        \multirow{1}*{NAM (ours)}	&	\bf 28.8	&	90.4	&	61.9	&	\bf 73.5	\\
        \bottomrule
    \end{tabular}
\end{table}

\begin{figure*}[t]

\begin{minipage}[t]{.24\linewidth}
  \centering
  \centerline{\includegraphics[width=3.8cm]{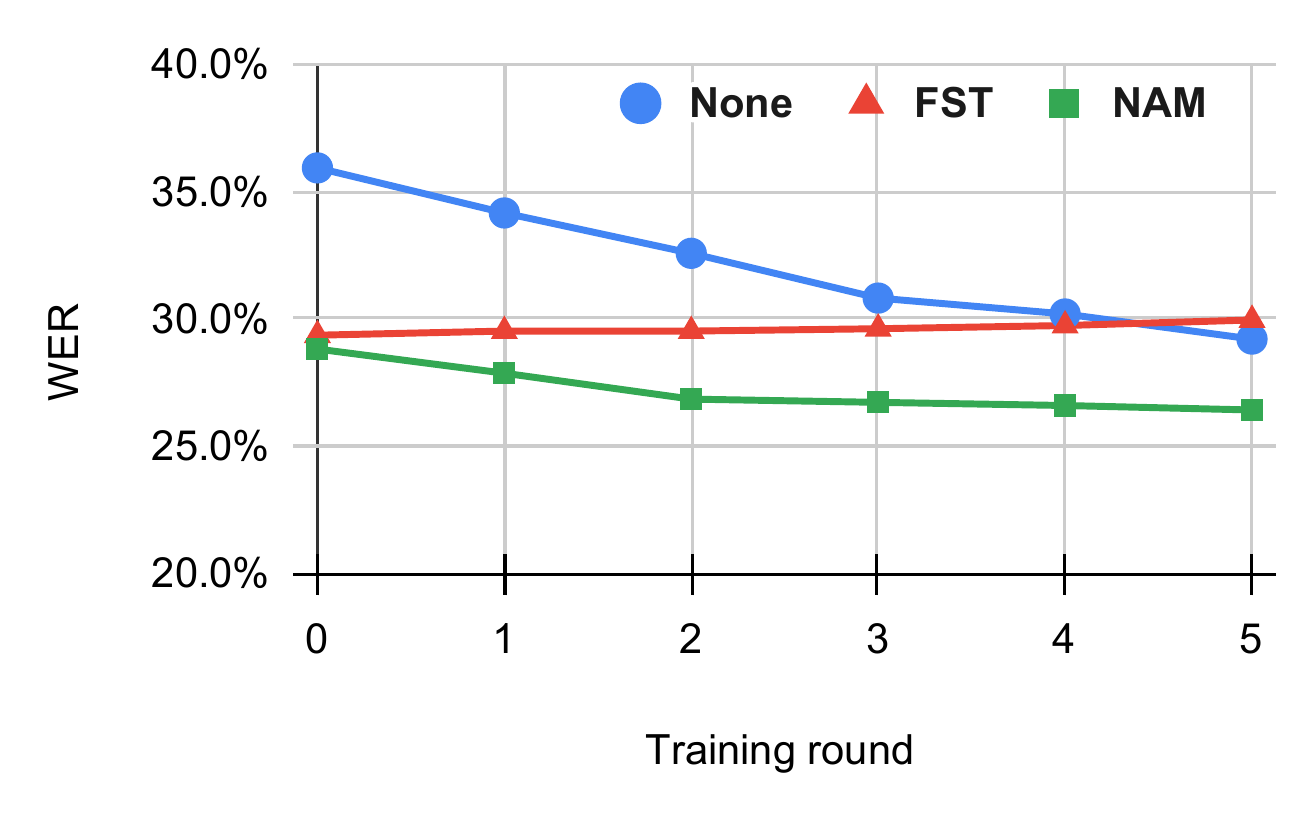}}
\end{minipage}
\begin{minipage}[t]{0.24\linewidth}
  \centering
  \centerline{\includegraphics[width=4.0cm]{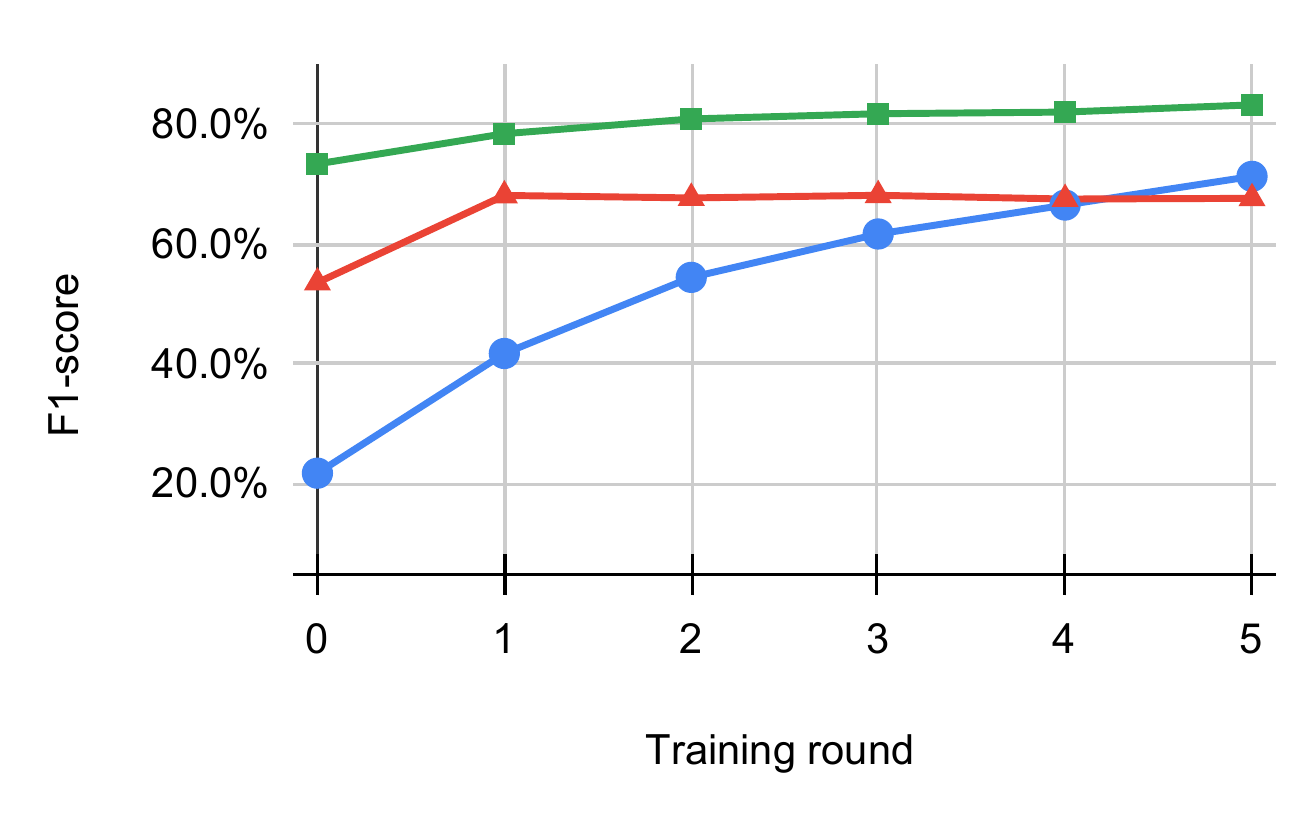}}
\end{minipage}
\begin{minipage}[t]{.24\linewidth}
  \centering
  \centerline{\includegraphics[width=4.0cm]{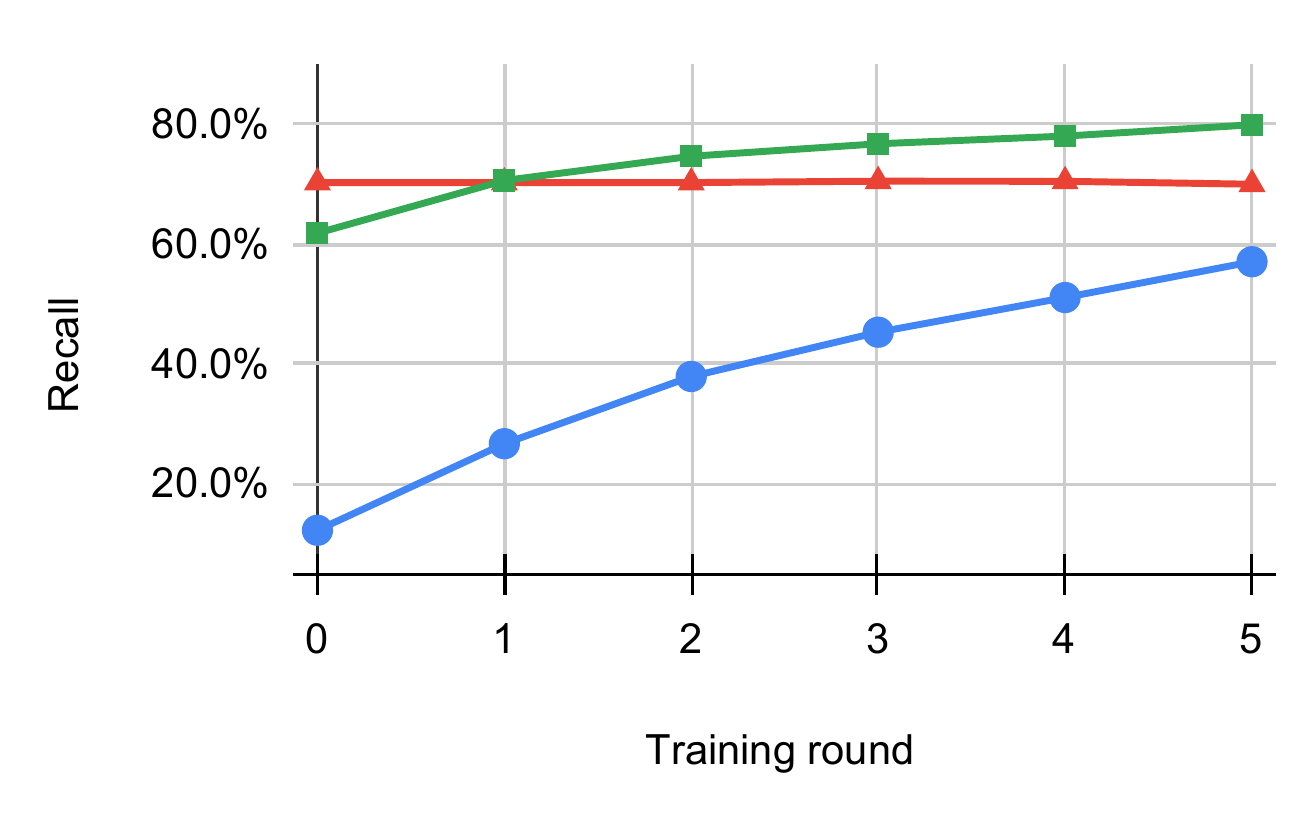}}
\end{minipage}
\begin{minipage}[t]{0.24\linewidth}
  \centering
  \centerline{\includegraphics[width=3.8cm]{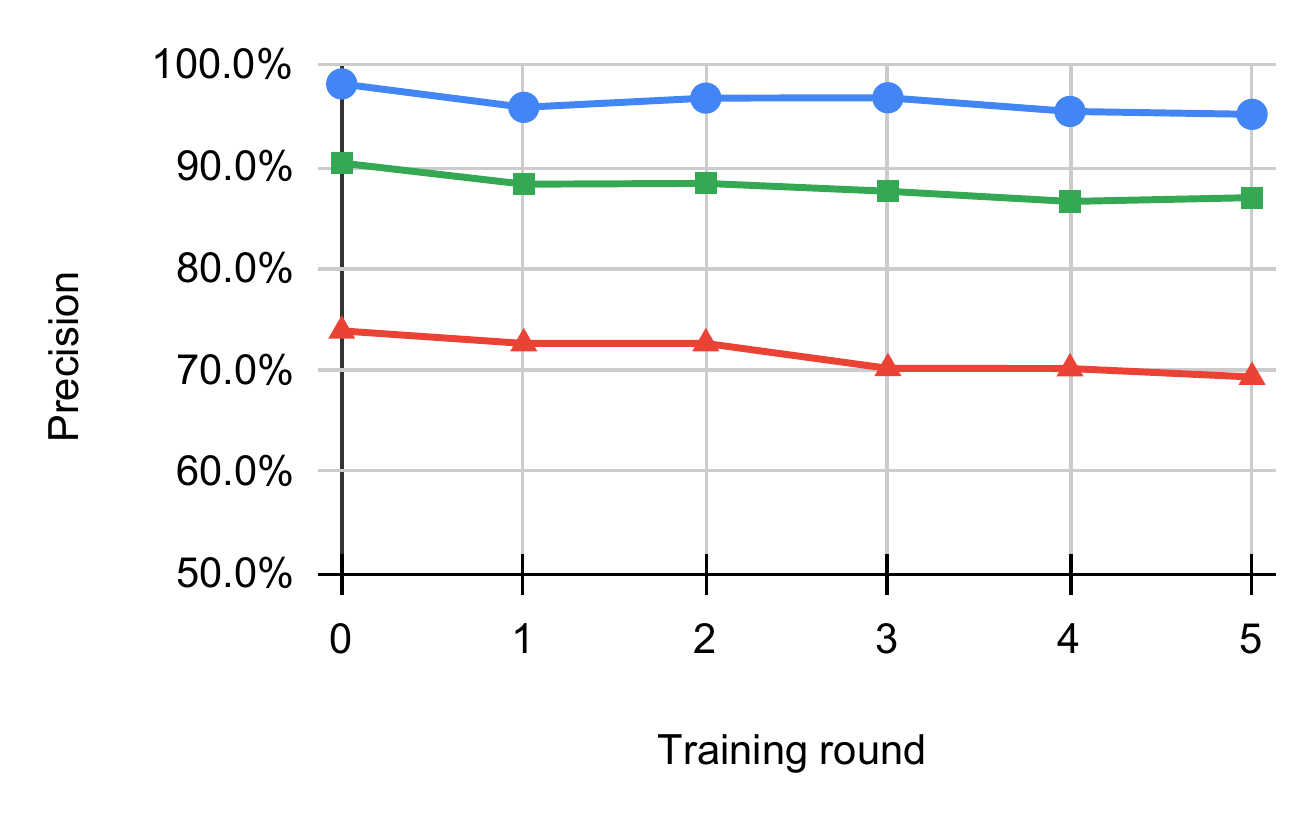}}
\end{minipage}
\caption{Comparison of the test metrics with increasing training rounds.}
\label{fig:personalization}
\end{figure*}

\subsection{Datasets}
\label{sec:dataset}
We train on many types of speech data~\cite{narayanan2018toward}. The data consists of anonymized and transcribed English utterances from different domains including accented speech, telephony speech, and YouTube. The total amount of training data is around 400k hours. We followed a similar data augmentation procedure as in~\cite{narayanan2021cascaded}. During training, we followed the procedure described in Section~\ref{sec:training} and annotated the transcript.

We run tests on the {\em Wiki-Names} corpus~\cite{sim19asru} to evaluate the proposed approach in an on-device simulation environment and compare with baseline models. The {\em Wiki-Names} corpus was designed for entity mention personalization. This data set contains read speech collected from 100 speakers. The text prompts used for the data collection were extracted from US English Wikipedia pages containing named entity mentions with type \textit{person} that are misrecognized by a reference ASR model. In this set, each speaker has 50 train, 10 dev and 10 test utterances that cover 5 entity mentions total per speaker. We have around 4.6 minutes of training data per speaker. Unless specified differently, we add 5 contextual phrases per utterance, one of which is in the utterance transcript and 4 distractor phrases.

\subsection{Model}
\label{sec:models}
The ASR model used in our large scale training on the multi-domain data and the on-device simulation experiments is an RNN-T model with Conformer~\cite{gulati2020conformer} encoder. This model is similar to those described in~\cite{narayanan2021cascaded}. The are 17 conformer blocks followed by the contextual biasing module.
128-dimensional log-Mel features are computed on a 32
milliseconds window, with 22 milliseconds overlap between neighboring windows. 4 subsequent feature vectors are stacked yielding a 512-dimensional input features, which is then sub-sampled by a factor of 3 across the time dimension. We used a WP model with vocabulary size of 4096 for the transcript and biasing text tokenization.

For the contextual biasing module, we used 5 Transformer blocks for context encoder and the input and hidden dimensions for the MHA with 8 heads were set to 512 and 1024, respectively. Contextual phrases are 1 to 3 words long, and the insertion probability p was 0.7 during training. The models were trained on v3-128 TPU Pod with a batch size of 4096 and Adam optimizer.



\begin{table}[t]
    \centering
    \caption{Results on {\em Wiki-Names} test set after personalized training. The best FST model (with 29.4\% WER) was selected based on the contextual adaptation results reported in Table~\ref{tab:adaptation}.}
    \label{tab:final_adapt}
    \begin{tabular}{c c c c c}
        \toprule
        \multirow{1}*{Model} & WER & Precision & Recall & F1-score \\
        \toprule
        \multirow{1}*{None}
        &    29.2    &  \bf 95.2    &    57.1    &    71.4  \\
        \multirow{1}*{FST}
        &   30.0 & 69.3 &	70.1 &	67.7\\
        \multirow{1}*{NAM (ours)}	&	\bf 26.4	&	87.0	&	\bf 80.0	&	\bf 83.4	\\
        \bottomrule
    \end{tabular}
\end{table}

\subsection{Fast Contextual Adaptation Results}
\label{sec:adaptation}
After training on the multi-domain data, we tested the models on the {\em Wiki-Names} corpus by providing entity mentions as biasing context to evaluate for fast adaptation behavior. The entity mentions are unseen during training and thus an efficient model must adapt on-the-fly to recognize the mentions correctly after provided with the biasing context. 

In addition to the proposed NAM model, we evaluate two other approaches: a simple non-contextual baseline model (None) and the same baseline model with the traditional FST biasing component~\cite{aleksic2015improved,mcgraw2016personalized}. We measure overall WER as well as entity mention specific Recall, Precision and F1-score metrics. For the FST biasing model, there are additional hyper-parameters that require properly tuned,  re-scoring weight value being the most important one. 

The results are summarized in Table~\ref{tab:adaptation}. For the FST baseline, we reported multiple results with different re-scoring weights. The best WER for the FST baseline is 29.4\% after tuning its re-scoring weight while the non-biasing baseline without any contextual biasing obtains 36.0\% WER. Our proposed approach achieves 28.8\%, which is 2\% relative WER improvement over the FST biasing models. 
As expected, the contextual biasing models obtained a better recall. However, the higher recall resulted in a decreasing precision. Overall our model outperforms the FST biasing baseline by 1.4\% F1-score. 

We also evaluated the same NAM model for an empty context scenario, where no biasing phrases are provided as input. Our model achieved 35.9\% WER, showing no regression error when no biasing context is available. We noted that the FST baseline is sensitive to the choice of re-scoring weight value and when not tuned properly, this can even worsen the performance. Tuning of this value is expensive as well, which our NAM model does not require.


\subsection{Continuous Personalization Results}
\label{sec:personalization}
We run an extensive evaluation in a continuous personalization scenario~\cite{sim21_interspeech} where the models are continually trained to further adapt for each speaker in {\em Wiki-Names}. We updated the RNN-T joint layers with 3.4 million parameters for each speaker. For memory efficiency, Adafactor optimization~\cite{shazeer2018adafactor} with a learning rate of 0.005 and without the first-order momentum was used as optimizer. The gradient clipping was set to 0.1. This setup closely simulates an on-device continuous personalization training.
The on-device training is performed with 2 epochs per training round and a mini-batch size of 10. We run total 5 training rounds. We used the approach described in~\cite{sim21_interspeech} to guard against catastrophic forgetting and model drift. 

Figure~\ref{fig:personalization} shows the detailed performance metrics on {\em Wiki-Names} test data as the models are continually trained from one round to another round and Table~\ref{tab:final_adapt} reports the final results after the continuous personalization. The WER for the non-biasing baseline and our proposed NAM models improve consistently throughout training and reaches 29.2\% and 26.4\% respectively. The recall and F1-score metrics are also improved. The NAM achieves 80.0\% recall and 83.4\% F1-score, outperforming the none and the FST biasing model by around 17\% and 23\% relative F1-score respectively.

However, for the FST baseline, the continuous personalization does not always improve the performance and it diverges only after the first round. The divergence is more severe in the case of the precision on the entity mentions. This is not surprising since the joint layer parameter updates during continuous training do not take into account the FST contextual biasing output. Also, once the model is changed due the parameter updates, the FST hyper-parameters, such as the re-scoring weight, need to be tuned again indicating that the re-scoring weight values poorly generalize to the new model. This is not suitable in an on-device personalization scenario. On the other hand, our proposed end-to-end approach offers a tighter integration with the ASR model and demonstrates robust improvement as the model is continuously personalized for each speaker.

\subsection{Model Ablation}
\label{sec:ablation}

\begin{figure}[t]
\begin{minipage}[b]{1.0\linewidth}
  \centering
  \centerline{\includegraphics[width=6.5cm]{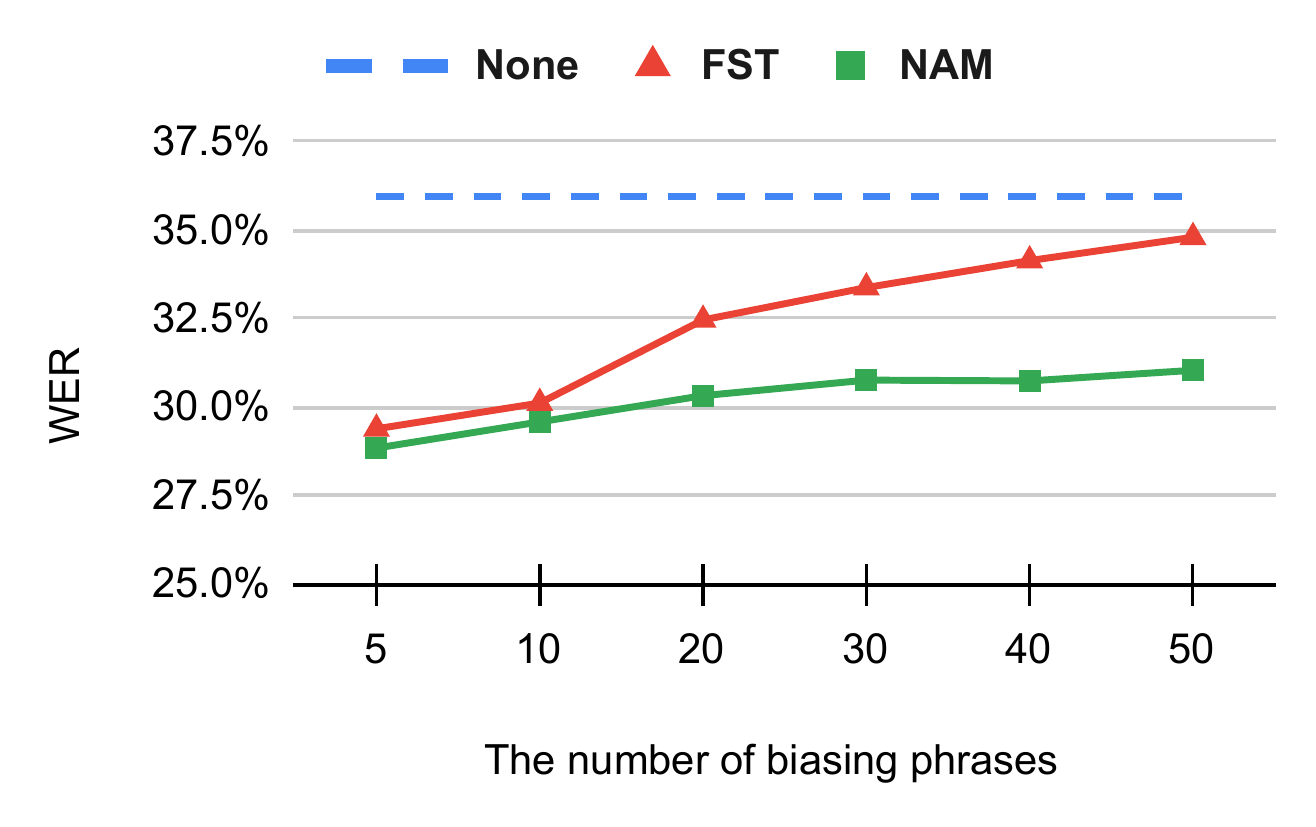}}
\end{minipage}
\caption{Comparison of FST and NAM on varying number of contextual phrases.}
\label{fig:fstvsnam}
\end{figure}

Finally, we performed ablations with varying number of contextual phrases as well as on model variants. We vary the context size from 10 to 50 entries consisting of only single positive and the remaining ant-context phrase.
Figure~\ref{fig:fstvsnam} compares the FST and NAM models. The FST WER increases significantly with a larger context size while the WER for our model-based end-to-end approach degrades gracefully. 

As for the model variants, we have 3 models including an adopted version of the previous CLAS model. They are listed in Table~\ref{tab:ablation} along with their performance. In the NAM-\textit{single} model, we construct only a single vector representation of each phrase whereas the NAM-\textit{left shift} does not apply the left-shifting for the value vectors. For the CLAS-\textit{encoder} model, we moved CLAS's contextual module to the top of the encoder and fed in the audio features. Learning a single representation for each contextual phrase hurts the performance more than the sharing the key and value vector without the left-shift, which underperforms the main NAM, specially in terms of recall. The CLAS-\textit{encoder} performs better than the non-biasing model, although its results are still far from that of the NAM model.

\begin{table}[t]
    \centering
    \caption{Results of model variants on {\em Wiki-Names} test set.}
    \label{tab:ablation}
    \begin{tabular}{c c c c c}
        \toprule
        \multirow{1}*{Model} & WER & Precision & Recall & F1-score \\
        \toprule
        \multirow{1}*{NAM} &
        28.8	&	90.4	&	61.9	&	73.5 \\
        \multirow{1}*{NAM-\textit{left shift}} & 
        29.6	&	90.9	&	56.5	&	69.7 \\
        \multirow{1}*{NAM-\textit{single}} & 
        30.5	&	90.3	&	50.5	&	64.7 \\
        \multirow{1}*{CLAS-\textit{encoder}} & 
        34.9	&	96.8	&	20.9	&	34.4 \\
        \bottomrule
    \end{tabular}
\end{table}

\section{Conclusion}
\label{sec:conclusion}
We introduced a novel end-to-end contextual biasing approach for fast adaptation of ASR models. We proposed a neural associative memory to store and retrieve fine-grained transitional information of each contextual phrase. 
Our approach is decoder-agnostic as it learns contextualized representations of audio encoder features by leveraging multi-modal learning.
The proposed model achieved a better initial as well as a personalized performance over strong baselines. When evaluated on increasing number of contextual phrases, our NAM model outperformed the FST biasing approach by up to 10.9\% relative WER.

\vfill\pagebreak

\bibliographystyle{IEEEbib}
\bibliography{refs}

\end{document}